\documentclass[showpacs,aps,graphicx]{revtex4}
\usepackage{graphicx}
\begin{document}

\title{Multipartite entanglement purification for three-level trapped atom systems}

\author{Yu-Bo Sheng,$^{1,2}$\footnote{Email address:
shengyb@njupt.edu.cn} Jiong Liu$^{1,2}$,  Sheng-Yang Zhao$^{1,2}$ }

\address{ $^1$Institute of Signal Processing  Transmission, Nanjing
University of Posts and Telecommunications, Nanjing, 210003,  China\\
$^2$Key Lab of Broadband Wireless Communication and Sensor Network
 Technology,Nanjing University of Posts and Telecommunications, Ministry of
 Education, Nanjing, 210003,
 China\\}
\begin{abstract}
We describe an entanglement purification protocol (EPP) for multipartite  three-level atomic entangled pair using photonic Faraday rotation. In this EPP, the multipartite atomic entangled state can be purified with the help of some single photons.
This protocol resorts to the
photonic Faraday rotation to realize the function of the controlled-Not (CNOT) gate.  The purified multipartite atomic entangled state can be retained and to be repeated to reach a higher fidelity.
\end{abstract}
\maketitle

\section{Introduction}
In the recent years, there have been substantial interesting applications in quantum information processing
(QIP) based on quantum entanglement \cite{rmp}. Many quantum information branches such as quantum teleportation \cite{teleportation,teleportation1},
quantum key distribution (QKD) \cite{Ekert91,QKDdeng1,QKDdeng2}, quantum secret sharing \cite{QSS1,QSS2,QSS3},quantum state sharing \cite{quantusatatesharing1,quantusatatesharing2,quantusatatesharing3}, quantum direction communication \cite{QSDC1,QSDC2,QSDC3} all need the quantum entanglement.
Especially, the multipartite entangled systems have many vital applications in quantum communication
and quantum computation, such as controlled teleportation \cite{teleportation1}, quantum state sharing \cite{QSS1,QSS2,QSS3},   and so on. All these tasks are based on the fact that the maximally entangled quantum channel with multipartite
entangled states shared by the legitimate distant participants has been set up beforehand. Unfortunately, the multipartite
entangled states will interact with the environment and suffer from the noise when  they are distributed to distant locations.
The quality of the entanglement will be decreased. The low quality of the entanglement will  decrease the fidelity
of quantum teleportation and quantum key distribution. It also  makes the quantum communication insecure.

Entanglement purification is one of the powerful way to extract some high-fidelity entangled states from a set of less-entangled systems \cite{Bennett,Deutsch,pan1,pan2,simon,shengpra1,shengpra2,shengpra3,lipra,caozl,fengpra,wangc3,wangc4}.
In 1996, Bennett \emph{et al.} proposed  an entanglement purification protocol (EPP) with controlled-not (CNOT) gate \cite{Bennett}. This protocol was improved
by Deustch \emph{et al.} in the same year \cite{Deutsch}. In 2001, Pan \emph{et al.} proposed an EPP with linear optics \cite{pan1,pan2}. In 2002, Simon and Pan  presented
an EPP with a currently available parametric down-conversion (PDC) source \cite{simon}. In 2008, the EPP based on the PDC source and cross-Kerr nonlinearity
was proposed \cite{shengpra1}. In 2010, the deterministic EPP with hyperentanglement was proposed \cite{shengpra2}. This EPP was improved with linear optics by Sheng and
Li, respectively \cite{shengpra3,lipra}. These EPPs are all focused on the bipartite entangled photon systems and there are only
several multipartite entanglement purification protocols (MEPPs). In 1998, Murao \emph{et al.} described an MEPP with CNOT gate for purifying the
Werner-type state \cite{murao}. Cheong \emph{et al.} extended this idea to purify the high-dimensional
multipartite quantum systems in 2007 \cite{highdemension}. In 2009, Sheng \emph{et al.} described an MEPP for optical systems with the help of
cross-Kerr nonlinearity \cite{shengepjd}. In 2011, Deng \emph{et al.} developed an high efficiency MEPP with the entanglement link from a subspace \cite{dengpra}.
Recently, the MEPPs for conduction electron systems and the multipartite electronic systems resorting the
quantum-dot and microcavities were proposed \cite{shengpla,dengqip,wangqip}.

Currently, trapped atoms with cavities which so called cavity quantum electrodynamics (QED) have become one of the promising candidates for QIP \cite{rmpqed}.
The theoretical and experimental works have been done, such as the conditional logic gating \cite{logicgate}, efficient generation of single photons \cite{singlephoton}, and so on.
The atoms trapped in the different distant local cavities with strongly interaction can be viewed as the quantum nodes in the quantum repeaters. The
 photons flying between these different quantum nodes as the quantum bus can be used to set up the quantum network for largely scaling the QIP.   The key point is that the single-photon input and output a low-Q cavity can moderate and couple with the trapped atom, which is so called
 microtoroidal resonator. Some theoretical works showed  that the polarized photons can obtain different phase shift after they interacting
with an atom trapped in a low-Q cavity. The phase shift can reach $\pi$ by choosing the appropriate parameters. It is named  the optical Faraday rotation \cite{faradayrotation}.
Based on the Faraday rotation, An \emph{et al.} discussed the quantum-information processing with a single photon by an input-output process with low-Q cavities \cite{fengmang1,fengmang2}.
Chen and Feng investigated the quantum gating on neutral atoms in low-Q cavities \cite{entanglementgeneration}. The quantum teleportation \cite{fengmang3}, controlled
teleportation \cite{cteleportationqip}, entanglement swapping \cite{swapping}  and entanglement concentration  for atoms were proposed \cite{atomconcentration}.
Motivated by the optical Faraday rotation, we  present an MEPP for trapped atoms in low-Q cavities. In this MEPP, the ancillary photons
and cavity systems play the role of the CNOT gate. This paper is organized as follows: In Sec. 2, we briefly explain the
basic principle of optical Faraday rotation. In Sec. 3, we describe the MEPP for three-atom Greenberger-Horne-Zeilinger (GHZ) state for a bit-flip error. In Sce. 4,
we describe the MEPP for a phase-flip error. In Sec. 5, we briefly explain the MEPP for $N$-atom systems. Finally, in Sec. 6,
we make a discussion and conclusion.

\section{Hybrid CNOT gate using photonic Faraday rotation}
In this section, we will discuss the basic principle for photonic Faraday rotation.
As shown in Fig. 1, a three-level atom is trapped in a cavity. Under the Jaynes-Cumming model, we can write the
Hamiltonian \cite{fengmang3,cteleportationqip,swapping}
\begin{eqnarray}
H=\sum_{j=L,R}[\frac{\hbar\omega_{0}}{2}\sigma_{_{jz}}+\hbar\omega_{c}a_{j}^{\dagger}a_{j}+i\hbar g(a_{j}\sigma_{j+}-a_{j}^{\dagger}\sigma_{j-})],
\end{eqnarray}
where $a^{\dagger}$ and $a$ are the creation and annihilation operators of the cavity field with frequency $\omega_{c}$, respectively.
$\sigma_{z}$, $\sigma_{+}$ and $\sigma_{-}$ are inversion, raising and lowering operators of the three-level atom with frequency difference
$\omega_{0}$ between the two levels $|e\rangle$ and $|g\rangle$. $|e\rangle$ is the excited state and  $|g\rangle$ is the
 ground state. The ground state has   two degenerated ground states  $|g_{R}\rangle$ and $|g_{L}\rangle$
which couple with a right ($|R\rangle$) and a left ($|L\rangle$) polarization photon, respectively. Consider the single-photon pulse input an optical cavity, as shown in Fig. 1.
Usually, the single photon pulse can be written as $|\Phi_{p}\rangle=\int^{T}_{0}f(t)a^{\dagger}_{in}(t)dt|vac\rangle$. The $a^{\dagger}_{in}(t)$ is
the one-dimensional field operator, which satisfies the commutation relation $[a_{in}(t),a^{\dagger}_{in}(t')]=\delta(t-t')$.
$f(t)$ in the normalized pulse shape as a function of $t$.
\begin{figure}[!h]
\begin{center}
\includegraphics[width=8cm,angle=0]{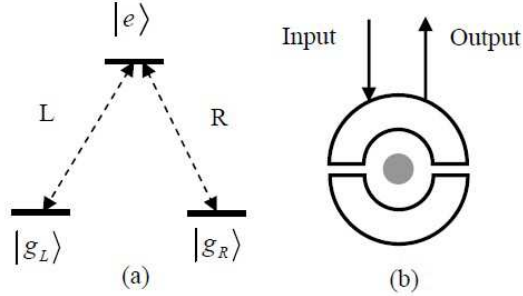}
\caption{(a) The relevant three-level atomic structure confined in a cavity. $|g_{R}\rangle$ and $|g_{L}\rangle$ are the two degenerated ground states
which couple with a right ($|R\rangle$) and a left ($|L\rangle$) polarization photon, respectively. (b) Atomic configuration of the three-level atom trapped in the low-Q cavity. Interaction between the atom and a single-photon pulse propagating input-output in the low-Q cavity according to the photonic Faraday rotation.}
\end{center}
\end{figure}
In the ration frame, the Heisenberg equation of the
operators $a_{j}$ and $\sigma_{j-}$ can be described as \cite{cteleportationqip}
\begin{eqnarray}
\dot{a}_{j}(t)&=&-[i(\omega_{c}-\omega_{p})+\frac{\kappa}{2}]a_{j}(t)-g\sigma_{j-}(t)-\sqrt{\kappa}a_{in,j}(t),\nonumber\\
\dot{\sigma}_{j-}(t)&=&-[i(\omega_{c}-\omega_{p})+\frac{\gamma}{2}]]\sigma_{j-}(t)-g\sigma_{iz}(t)a_{j}(t)+\sqrt{\gamma}\sigma_{z}(t)b_{in,j}(t).
\end{eqnarray}
The $\kappa$ is the cavity damping rate and $[b_{in}(t),b^{\dagger}_{in}(t')]=\delta(t-t')$.
The input and output fields of the cavity reads as
\begin{eqnarray}
a_{out,j}(t)=a_{in,j}(t)+\sqrt{\kappa}a_{j},
\end{eqnarray}
with $j=L,R$. Suppose that plus an adiabatic approximation, and $b_{in,j}\simeq0$, we can adiabatically eliminate
cavity mode and get the relation of the cavity mode as
\begin{eqnarray}
r_{j}(\omega_{p})=\frac{[i(\omega_{c}-\omega_{p})-\frac{\kappa}{2}][i(\omega_{0}-\omega_{p})+\frac{\gamma}{2}]
+g^{2}}{[i(\omega_{c}-\omega_{p})+\frac{\kappa}{2}][i(\omega_{0}-\omega_{p})+\frac{\gamma}{2}]+g^{2}}.\label{reflect1}
\end{eqnarray}
where $r(\omega)\equiv\frac{a_{out,j(t)}}{a_{in,j(t)}}$ is the reflection coefficient of the atom-cavity system. Interestingly,
from above equation, if $g=0$, we can obtain
\begin{eqnarray}
r_{j0}(\omega_{p})=\frac{i(\omega_{c}-\omega_{p})-\frac{\kappa}{2}}{i(\omega_{c}-\omega_{p})+\frac{\kappa}{2}}.
\end{eqnarray}
Due to the high damping rate of the cavity, the absolute value of $r_{j}(\omega_{p})$ can reach to unity.
It means that the photon experiences a very weak absorption and we only need to consider that
the output photon only experiences a pure phase shift $\theta$ with $r_{j}(\omega_{p})=e^{i\theta}$. In a special case that
the atom does not couple with the cavity or couples with an empty, we have $r_{j0}(\omega_{p})=e^{i\theta_{0}}$.
Therefore, the $\Theta_{F}=(\theta-\theta_{0})/2$ is called the Faraday rotation.
If $\omega_{c}=\omega_{0}$, $\omega_{p}=\omega_{c}-\frac{\kappa}{2}$ and $g=\frac{\kappa}{2}$, we can obtain $\theta=\pi$ and
$\theta_{0}=\frac{\pi}{2}$. In this way, the evolution of the atom and photon input and output the cavity can be described as\cite{atomconcentration}
\begin{eqnarray}
|L\rangle|g_{L}\rangle\rightarrow-|L\rangle|g_{L}\rangle, |R\rangle|g_{L}\rangle\rightarrow i|R\rangle|g_{L}\rangle,\nonumber\\
|L\rangle|g_{R}\rangle\rightarrow i|L\rangle|g_{R}\rangle, |R\rangle|g_{R}\rangle\rightarrow -|R\rangle|g_{R}\rangle.\label{relation}
\end{eqnarray}
If the photon passes through two cavities, one can get
\begin{eqnarray}
|L\rangle|g_{L}\rangle|g_{L}\rangle\rightarrow|L\rangle|g_{L}\rangle|g_{L}\rangle, |R\rangle|g_{L}\rangle|g_{L}\rangle\rightarrow -|R\rangle|g_{L}\rangle|g_{L}\rangle,\nonumber\\
|L\rangle|g_{R}\rangle|g_{R}\rangle\rightarrow -|L\rangle|g_{R}\rangle|g_{R}\rangle, |R\rangle|g_{R}\rangle|g_{R}\rangle\rightarrow |R\rangle|g_{R}\rangle|g_{R}\rangle,\nonumber\\
|L\rangle|g_{L}\rangle|g_{R}\rangle\rightarrow-i|L\rangle|g_{L}\rangle|g_{R}\rangle, |R\rangle|g_{L}\rangle|g_{R}\rangle\rightarrow -i|R\rangle|g_{L}\rangle|g_{L}\rangle,\nonumber\\
|L\rangle|g_{R}\rangle|g_{L}\rangle\rightarrow -i|L\rangle|g_{R}\rangle|g_{R}\rangle, |R\rangle|g_{R}\rangle|g_{L}\rangle\rightarrow -i|R\rangle|g_{R}\rangle|g_{L}\rangle.\label{relation1}
\end{eqnarray}
From Eq. (\ref{relation1}), if the initial state of photon is $\frac{1}{\sqrt{2}}(|L\rangle+|R\rangle)$, and the
 atomic state is $|g_{L}\rangle|g_{L}\rangle$ or $|g_{R}\rangle|g_{R}\rangle$, the photonic state will become $\frac{1}{\sqrt{2}}(|L\rangle-|R\rangle)$, while
 it does not change if the atomic state is  $|g_{L}\rangle|g_{R}\rangle$ or $|g_{R}\rangle|g_{L}\rangle$. Therefore, the polarization of the photon flipping  or not
 flipping is decided by the  states of two atoms. It  essentially  acts as the role of CNOT gate. Interestingly, the CNOT gate is between different
 qubit and different degrees of freedom, it is a hybrid CNOT gate.

 \section{MEPP for multipartite atomic entangled state for bit-flip error}
In this section, we will start to explain the basic principle of MEPP.
We first take the three-particle system in GHZ state as an example. A three-particle GHZ state can be described as
\begin{eqnarray}
|\Phi^{+}_{0}\rangle=\frac{1}{\sqrt{2}}(|g_{L}\rangle|g_{L}\rangle|g_{L}\rangle+|g_{R}\rangle|g_{R}\rangle|g_{R}\rangle).\label{GHZ1}
\end{eqnarray}
The other seven states of the GHZ states are written as
\begin{eqnarray}
|\Phi^{-}_{0}\rangle=\frac{1}{\sqrt{2}}(|g_{L}\rangle|g_{L}\rangle|g_{L}\rangle-|g_{R}\rangle|g_{R}\rangle|g_{R}\rangle),\nonumber\\
|\Phi^{\pm}_{1}\rangle=\frac{1}{\sqrt{2}}(|g_{L}\rangle|g_{L}\rangle|g_{R}\rangle\pm|g_{R}\rangle|g_{R}\rangle|g_{L}\rangle),\nonumber\\
|\Phi^{\pm}_{2}\rangle=\frac{1}{\sqrt{2}}(|g_{L}\rangle|g_{R}\rangle|g_{L}\rangle\pm|g_{R}\rangle|g_{L}\rangle|g_{R}\rangle),\nonumber\\
|\Phi^{\pm}_{3}\rangle=\frac{1}{\sqrt{2}}(|g_{R}\rangle|g_{L}\rangle|g_{L}\rangle\pm|g_{L}\rangle|g_{R}\rangle|g_{R}\rangle).\label{GHZ2}
\end{eqnarray}
Usually, in a distributed quantum network, if the initial state is $|\Phi^{+}_{0}\rangle$, it will couple with the noisy environment and  become other states with some probability.
For instance, if a bit-flip error occurs on the first qubit and it will become $|\Phi^{+}_{1}\rangle$. If a phase-flip error occurs, it will become
$|\Phi^{-}_{0}\rangle$. Certainly, both the bit-flip error and the phase-flip error will occur and lead the initial state $|\Phi^{+}_{0}\rangle$
become $|\Phi^{-}_{1}\rangle$, $|\Phi^{-}_{2}\rangle$ or $|\Phi^{-}_{3}\rangle$.
We start to discuss the MEPP with a simple example. Suppose a bit-flip error occurs on the first qubit with the probability of $1-F$.
The mixed state can be described as
\begin{eqnarray}
\rho_{B}=F|\Phi^{+}_{0}\rangle\langle\Phi^{+}_{0}|+(1-F)|\Phi^{+}_{1}\rangle\langle\Phi^{+}_{1}|.\label{mixed}
\end{eqnarray}
\begin{figure}[!h]
\begin{center}
\includegraphics[width=12cm,angle=0]{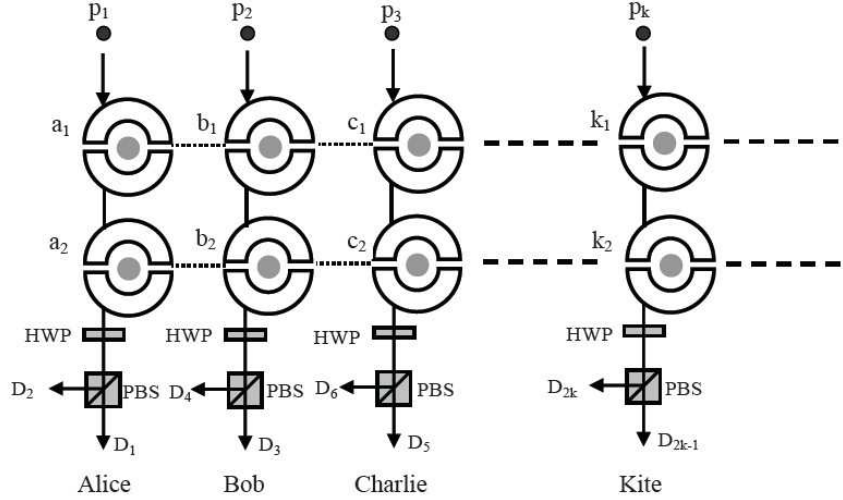}
\caption{Schematic diagram
showing the principle of this MEPP. Two pairs of low fidelity $N$-atom trapped in the cavities
are distributed to Alice, Bob, Charlie, $\cdots$, etc.. Each parties own
   two atoms. The one is from the first pair and the other is from the second pair.
   p$_{1}$, p$_{2}$,$\cdots$, p$_{n}$ are the auxiliary single photons, which will be detected by the single-photon detectors
D$_{1}$, D$_{2}$,$\cdots$, D$_{2k}$ to decide to retain or discard the atoms.}
\end{center}
\end{figure}
The subscript $B$ means a bit-flip error. From Fig. 2,  Alice, Bob and Charlie own two copies of the mixed state $\rho_{a_{1}b_{1}c_{1}}$ and
$\rho_{a_{2}b_{2}c_{2}}$ with the same form of Eq. (\ref{mixed}). Alice owns the atoms $a_{1}$ and $a_{2}$.
Bob owns the atoms $b_{1}$ and $b_{2}$ and Charlie owns $c_{1}$ and $c_{2}$, respectively.
The two pairs of mixed states can be viewed as four combinations of the pure states: with the probability
of $F^{2}$, it is in the state $|\Phi^{+}_{0}\rangle_{a_{1}b_{1}c_{1}}\otimes|\Phi^{+}_{0}\rangle_{a_{2}b_{2}c_{2}}$,
with the same probability of $F(1-F)$, it is in the state $|\Phi^{+}_{0}\rangle_{a_{1}b_{1}c_{1}}\otimes|\Phi^{+}_{1}\rangle_{a_{2}b_{2}c_{2}}$
or $|\Phi^{+}_{1}\rangle_{a_{1}b_{1}c_{1}}\otimes|\Phi^{+}_{0}\rangle_{a_{2}b_{2}c_{2}}$. With the probability of $(1-F)^{2}$, it is in the
state $|\Phi^{+}_{1}\rangle_{a_{1}b_{1}c_{1}}\otimes|\Phi^{+}_{1}\rangle_{a_{2}b_{2}c_{2}}$.
From Eqs. (\ref{GHZ1}) and (\ref{GHZ2}), $|\Phi^{+}_{0}\rangle_{a_{1}b_{1}c_{1}}\otimes|\Phi^{+}_{0}\rangle_{a_{2}b_{2}c_{2}}$ can be written as
\begin{eqnarray}
&&|\Phi^{+}_{0}\rangle_{a_{1}b_{1}c_{1}}\otimes|\Phi^{+}_{0}\rangle_{a_{2}b_{2}c_{2}}=\frac{1}{\sqrt{2}}(|g_{L}\rangle_{a_{1}}|g_{L}\rangle_{b_{1}}|g_{L}\rangle_{c_{1}}
+|g_{R}\rangle_{a_{1}}|g_{R}\rangle_{b_{1}}|g_{R}\rangle_{c_{1}})\nonumber\\
&\otimes&\frac{1}{\sqrt{2}}(|g_{L}\rangle_{a_{2}}|g_{L}\rangle_{b_{2}}|g_{L}\rangle_{c_{2}}+|g_{R}\rangle_{a_{2}}|g_{R}\rangle_{b_{2}}|g_{R}\rangle_{c_{2}})\nonumber\\
&=&\frac{1}{2}(|g_{L}\rangle_{a_{1}}|g_{L}\rangle_{a_{2}}|g_{L}\rangle_{b_{1}}|g_{L}\rangle_{b_{2}}|g_{L}\rangle_{c_{1}}|g_{L}\rangle_{c_{2}}\nonumber\\
&+&|g_{R}\rangle_{a_{1}}|g_{R}\rangle_{a_{2}}|g_{R}\rangle_{b_{1}}|g_{R}\rangle_{b_{2}}|g_{R}\rangle_{c_{1}}|g_{R}\rangle_{c_{2}}
+|g_{L}\rangle_{a_{1}}|g_{R}\rangle_{a_{2}}|g_{L}\rangle_{b_{1}}|g_{R}\rangle_{b_{2}}|g_{L}\rangle_{c_{1}}|g_{R}\rangle_{c_{2}}\nonumber\\
&+&|g_{R}\rangle_{a_{1}}|g_{L}\rangle_{a_{2}}|g_{R}\rangle_{b_{1}}|g_{L}\rangle_{b_{2}}|g_{R}\rangle_{c_{1}}|g_{L}\rangle_{c_{2}}).\label{cross1}
\end{eqnarray}
The cross-combination item $|\Phi^{+}_{0}\rangle_{a_{1}b_{1}c_{1}}\otimes|\Phi^{+}_{1}\rangle_{a_{2}b_{2}c_{2}}$ can be written as
\begin{eqnarray}
&&|\Phi^{+}_{0}\rangle_{a_{1}b_{1}c_{1}}\otimes|\Phi^{+}_{1}\rangle_{a_{2}b_{2}c_{2}}=\frac{1}{\sqrt{2}}(|g_{L}\rangle_{a_{1}}|g_{L}\rangle_{b_{1}}|g_{L}\rangle_{c_{1}}
+|g_{R}\rangle_{a_{1}}|g_{R}\rangle_{b_{1}}|g_{R}\rangle_{c_{1}})\nonumber\\
 &\otimes&\frac{1}{\sqrt{2}}(|g_{L}\rangle_{a_{2}}|g_{L}\rangle_{b_{2}}|g_{R}\rangle_{c_{2}}
+|g_{R}\rangle_{a_{2}}|g_{R}\rangle_{b_{2}}|g_{L}\rangle_{c_{2}})\nonumber\\
&=&\frac{1}{2}(|g_{L}\rangle_{a_{1}}|g_{L}\rangle_{a_{2}}|g_{L}\rangle_{b_{1}}|g_{L}\rangle_{b_{2}}|g_{L}\rangle_{c_{1}}|g_{R}\rangle_{c_{2}}\nonumber\\
&+&|g_{R}\rangle_{a_{1}}|g_{R}\rangle_{a_{2}}|g_{R}\rangle_{b_{1}}|g_{R}\rangle_{b_{2}}|g_{R}\rangle_{c_{1}}|g_{L}\rangle_{c_{2}}
+|g_{L}\rangle_{a_{1}}|g_{R}\rangle_{a_{2}}|g_{L}\rangle_{b_{1}}|g_{R}\rangle_{b_{2}}|g_{L}\rangle_{c_{1}}|g_{L}\rangle_{c_{2}}\nonumber\\
&+&|g_{R}\rangle_{a_{1}}|g_{L}\rangle_{a_{2}}|g_{R}\rangle_{b_{1}}|g_{L}\rangle_{b_{2}}|g_{R}\rangle_{c_{1}}|g_{R}\rangle_{c_{2}}).\label{cross2}
\end{eqnarray}
One can also get
\begin{eqnarray}
&&|\Phi^{+}_{1}\rangle_{a_{1}b_{1}c_{1}}\otimes|\Phi^{+}_{0}\rangle_{a_{2}b_{2}c_{2}}
=\frac{1}{2}(|g_{L}\rangle_{a_{1}}|g_{L}\rangle_{a_{2}}|g_{L}\rangle_{b_{1}}|g_{L}\rangle_{b_{2}}|g_{R}\rangle_{c_{1}}|g_{L}\rangle_{c_{2}}\nonumber\\
&+&|g_{R}\rangle_{a_{1}}|g_{R}\rangle_{a_{2}}|g_{R}\rangle_{b_{1}}|g_{R}\rangle_{b_{2}}|g_{L}\rangle_{c_{1}}|g_{R}\rangle_{c_{2}}
+|g_{L}\rangle_{a_{1}}|g_{R}\rangle_{a_{2}}|g_{L}\rangle_{b_{1}}|g_{R}\rangle_{b_{2}}|g_{R}\rangle_{c_{1}}|g_{R}\rangle_{c_{2}}\nonumber\\
&+&|g_{R}\rangle_{a_{1}}|g_{L}\rangle_{a_{2}}|g_{R}\rangle_{b_{1}}|g_{L}\rangle_{b_{2}}|g_{L}\rangle_{c_{1}}|g_{L}\rangle_{c_{2}}),\label{cross3}
\end{eqnarray}
and
\begin{eqnarray}
&&|\Phi^{+}_{1}\rangle_{a_{1}b_{1}c_{1}}\otimes|\Phi^{+}_{1}\rangle_{a_{2}b_{2}c_{2}}
=\frac{1}{2}(|g_{L}\rangle_{a_{1}}|g_{L}\rangle_{a_{2}}|g_{L}\rangle_{b_{1}}|g_{L}\rangle_{b_{2}}|g_{R}\rangle_{c_{1}}|g_{R}\rangle_{c_{2}}\nonumber\\
&+&|g_{R}\rangle_{a_{1}}|g_{R}\rangle_{a_{2}}|g_{R}\rangle_{b_{1}}|g_{R}\rangle_{b_{2}}|g_{L}\rangle_{c_{1}}|g_{L}\rangle_{c_{2}}
+|g_{L}\rangle_{a_{1}}|g_{R}\rangle_{a_{2}}|g_{L}\rangle_{b_{1}}|g_{R}\rangle_{b_{2}}|g_{R}\rangle_{c_{1}}|g_{L}\rangle_{c_{2}}\nonumber\\
&+&|g_{R}\rangle_{a_{1}}|g_{L}\rangle_{a_{2}}|g_{R}\rangle_{b_{1}}|g_{L}\rangle_{b_{2}}|g_{L}\rangle_{c_{1}}|g_{R}\rangle_{c_{2}}).\label{cross4}
\end{eqnarray}
Alice, Bob and Charlie first prepare the same polarized single photons of the form $\frac{1}{\sqrt{2}}(|L\rangle+|R\rangle)$, named p$_{1}$,
p$_{2}$, and p$_{3}$,  and
then let them entrance into each cavities, respectively.  After the three photons all passing through the cavities and they will be detected
by the single-photon detectors. From Eq. (\ref{relation1}), the polarization of the photon will be flipped if the states of the
two atoms are different. From Eqs. (\ref{cross1}) to (\ref{cross4}), the cross-combinations items $|\Phi^{+}_{0}\rangle_{a_{1}b_{1}c_{1}}\otimes|\Phi^{+}_{1}\rangle_{a_{2}b_{2}c_{2}}$ and $|\Phi^{+}_{1}\rangle_{a_{1}b_{1}c_{1}}\otimes|\Phi^{+}_{0}\rangle_{a_{2}b_{2}c_{2}}$
 cannot lead all the
polarization of the photons same. For example, the items $|g_{L}\rangle_{a_{1}}|g_{L}\rangle_{a_{2}}|g_{L}\rangle_{b_{1}}|g_{L}\rangle_{b_{2}}|g_{L}\rangle_{c_{1}}|g_{R}\rangle_{c_{2}}$ and $
|g_{R}\rangle_{a_{1}}|g_{R}\rangle_{a_{2}}|g_{R}\rangle_{b_{1}}|g_{R}\rangle_{b_{2}}|g_{R}\rangle_{c_{1}}|g_{L}\rangle_{c_{2}}$
in  $|\Phi^{+}_{0}\rangle_{a_{1}b_{1}c_{1}}\otimes|\Phi^{+}_{1}\rangle_{a_{2}b_{2}c_{2}}$ will lead the photons in Alice and
Bob flip, while the polarization of photon in Charlie does not change. On the other hand, the items $|g_{L}\rangle_{a_{1}}|g_{R}\rangle_{a_{2}}|g_{L}\rangle_{b_{1}}|g_{R}\rangle_{b_{2}}|g_{L}\rangle_{c_{1}}|g_{L}\rangle_{c_{2}}$ and
$|g_{R}\rangle_{a_{1}}|g_{L}\rangle_{a_{2}}|g_{R}\rangle_{b_{1}}|g_{L}\rangle_{b_{2}}|g_{R}\rangle_{c_{1}}|g_{R}\rangle_{c_{2}}$ will lead the
photons in Alice and Bob do not change while the polarization of photon in Charlie will change. Therefore, by selecting the cases that
all the polarization of the photons are the same, they will eliminate the contribution of the cross-combination items.
In Fig. 2, the have-wave plate (HWP) can make $\frac{1}{\sqrt{2}}(|L\rangle+|R\rangle)\rightarrow|L\rangle$ and
$\frac{1}{\sqrt{2}}(|L\rangle-|R\rangle)\rightarrow|R\rangle$. The polarization beam splitter (PBS) can transmit the
$|L\rangle$ polarization photon and reflect the $|R\rangle$ polarization photon. Finally, if the detectors D$_{1}$D$_{3}$D$_{5}$ fire,
the original state will become
\begin{eqnarray}
|\Phi_{0}\rangle&=&\frac{1}{\sqrt{2}}(|g_{L}\rangle_{a_{1}}|g_{L}\rangle_{b_{1}}|g_{L}\rangle_{c_{1}}|g_{L}\rangle_{a_{2}}|g_{L}\rangle_{b_{2}}|g_{L}\rangle_{c_{2}}\nonumber\\
&+&|g_{R}\rangle_{a_{1}}|g_{R}\rangle_{c_{1}}|g_{R}\rangle_{b_{1}}|g_{R}\rangle_{b_{2}}|g_{R}\rangle_{a_{2}}|g_{R}\rangle_{c_{2}}),
\end{eqnarray}
with the probability of $\frac{F^{2}}{2}$, and it will become
\begin{eqnarray}
|\Phi_{1}\rangle&=&\frac{1}{\sqrt{2}}(|g_{L}\rangle_{a_{1}}|g_{L}\rangle_{b_{1}}|g_{R}\rangle_{c_{1}}|g_{L}\rangle_{a_{2}}|g_{L}\rangle_{b_{2}}|g_{R}\rangle_{c_{2}}\nonumber\\
&+&|g_{R}\rangle_{a_{1}}|g_{R}\rangle_{b_{1}}|g_{L}\rangle_{c_{1}}|g_{R}\rangle_{a_{2}}|g_{R}\rangle_{b_{2}}|g_{L}\rangle_{c_{2}}),
\end{eqnarray}
with the probability of $\frac{(1-F)^{2}}{2}$.
Then they all perform the Hadamard operations on the atoms $a_{2}$, $b_{2}$, and $c_{2}$. The Hadamard operation will make
$|g_{L}\rangle\rightarrow\frac{1}{\sqrt{2}}(|g_{L}\rangle+|g_{R}\rangle)$ and $|g_{R}\rangle\rightarrow\frac{1}{\sqrt{2}}(|g_{L}\rangle-|g_{R}\rangle)$.
Then the $|\Phi_{0}\rangle$ becomes
\begin{eqnarray}
|\Phi_{0}\rangle&\rightarrow&\frac{1}{\sqrt{2}}(|g_{L}\rangle_{a_{1}}|g_{L}\rangle_{b_{1}}|g_{L}\rangle_{c_{1}}(\frac{1}{\sqrt{2}})^{\otimes3}(|g_{L}\rangle+|g_{R}\rangle)^{\otimes3}\nonumber\\
&+&|g_{R}\rangle_{a_{1}}|g_{R}\rangle_{b_{1}}|g_{R}\rangle_{c_{1}}(\frac{1}{\sqrt{2}})^{\otimes3}(|g_{L}\rangle-|g_{R}\rangle)^{\otimes3}).
\end{eqnarray}
The $|\Phi_{1}\rangle$ becomes
\begin{eqnarray}
|\Phi_{1}\rangle&\rightarrow&\frac{1}{\sqrt{2}}(|g_{L}\rangle_{a_{1}}|g_{L}\rangle_{b_{1}}|g_{R}\rangle_{c_{1}}(\frac{1}{\sqrt{2}})^{\otimes3}(|g_{L}\rangle+|g_{R}\rangle)^{\otimes2}(|g_{L}\rangle-|g_{R}\rangle)\nonumber\\ &+&|g_{R}\rangle_{a_{1}}|g_{R}\rangle_{b_{1}}|g_{L}\rangle_{c_{1}}(\frac{1}{\sqrt{2}})^{\otimes3}(|g_{L}\rangle-|g_{R}\rangle)^{\otimes}((|g_{L}\rangle+|g_{R}\rangle)).
\end{eqnarray}
In above two equations, we omit the subscripts $a_{2}b_{2}c_{2}$ for simple. Alice, Bob and Charlie measure
the atoms $a_{2}b_{2}c_{2}$ in the basis $\{|g_{L}\rangle,|g_{R}\rangle\}$. If the measurement result is $|g_{L}\rangle|g_{L}\rangle|g_{L}\rangle$, $|g_{L}\rangle|g_{R}\rangle|g_{R}\rangle$,$|g_{R}\rangle|g_{L}\rangle|g_{R}\rangle$, or $|g_{R}\rangle|g_{R}\rangle|g_{L}\rangle$, they
will obtain the $|\Phi^{+}_{0}\rangle_{a_{1}b_{1}c_{1}}$ with the fidelity $F'=\frac{F^{2}}{F^{2}+(1-F)^{2}}$.
If $F>\frac{1}{2}$, then $F'>F$. If the measurement result on atoms $a_{2}b_{2}c_{2}$  is  $|g_{L}\rangle|g_{L}\rangle|g_{R}\rangle$, $|g_{L}\rangle|g_{R}\rangle|g_{L}\rangle$, $|g_{R}\rangle|g_{L}\rangle|g_{L}\rangle$, or $|g_{R}\rangle|g_{R}\rangle|g_{R}\rangle$,
they will obtain the $|\Phi^{-}_{0}\rangle_{a_{1}b_{1}c_{1}}$ with the same fidelity $F'$. In this way, they only need to perform a phase-flip operation
on one of their atoms to obtain $|\Phi^{+}_{0}\rangle_{a_{1}b_{1}c_{1}}$.

Certainly, if the single-photon detectors  D$_{2}$D$_{4}$D$_{6}$ fire, they will obtain
\begin{eqnarray}
|\Phi_{0}\rangle'&=&\frac{1}{\sqrt{2}}(|g_{L}\rangle_{a_{1}}|g_{L}\rangle_{b_{1}}|g_{L}\rangle_{c_{1}}|g_{R}\rangle_{a_{2}}|g_{R}\rangle_{b_{2}}|g_{R}\rangle_{c_{2}}\nonumber\\
&+&|g_{R}\rangle_{a_{1}}|g_{R}\rangle_{c_{1}}|g_{R}\rangle_{b_{1}}|g_{L}\rangle_{b_{2}}|g_{L}\rangle_{a_{2}}|g_{L}\rangle_{c_{2}}),
\end{eqnarray}
with the probability of $\frac{F^{2}}{2}$,
and obtain
\begin{eqnarray}
|\Phi_{1}\rangle'&=&\frac{1}{\sqrt{2}}(|g_{L}\rangle_{a_{1}}|g_{L}\rangle_{b_{1}}|g_{R}\rangle_{c_{1}}|g_{R}\rangle_{a_{2}}|g_{R}\rangle_{b_{2}}|g_{L}\rangle_{c_{2}}\nonumber\\
&+&|g_{R}\rangle_{a_{1}}|g_{R}\rangle_{b_{1}}|g_{L}\rangle_{c_{1}}|g_{L}\rangle_{a_{2}}|g_{L}\rangle_{b_{2}}|g_{R}\rangle_{c_{2}}),
\end{eqnarray}
with the probability of $\frac{(1-F)^{2}}{2}$.
In this case, they first perform  bit-flip operations on the atoms $a_{2}$, $b_{2}$, and $c_{2}$, which makes $|\Phi_{0}\rangle'$ become
 $|\Phi_{0}\rangle$ and $|\Phi_{1}\rangle'$ becomes $|\Phi_{1}\rangle$. In this way, they can obtain the same result as described above.

In above description, we explained the purifying the bit-flip error on the third qubit, which makes   $|\Phi_{0}^{+}\rangle$  become
   $|\Phi_{1}^{+}\rangle$. In a practical environment, the bit-flip error may occur on  all the qubits, and leads the
   original pure maximally entangled state  $|\Phi_{0}^{+}\rangle$  become
   \begin{eqnarray}
   \rho'=F_{0}|\Phi^{+}_{0}\rangle\langle\Phi^{+}_{0}|+F_{1}|\Phi^{+}_{1}\rangle\langle\Phi^{+}_{1}|
   +F_{2}|\Phi^{+}_{2}\rangle\langle\Phi^{+}_{2}|+|F_{3}|\Phi^{+}_{3}\rangle\langle\Phi^{+}_{3}|,\label{mixed2}
 \end{eqnarray}
 with $F_{0}+F_{1}+F_{2}+F_{3}=1$.
 Using the same principle described above, by choosing the same polarization of the three photons, they will obtain
 a new mixed state of the form
 \begin{eqnarray}
   \rho''=F'_{0}|\Phi^{+}_{0}\rangle\langle\Phi^{+}_{0}|+F'_{1}|\Phi^{+}_{1}\rangle\langle\Phi^{+}_{1}|
   +F'_{2}|\Phi^{+}_{2}\rangle\langle\Phi^{+}_{2}|+F'_{3}|\Phi^{+}_{3}\rangle\langle\Phi^{+}_{3}|,\label{mixed3}
 \end{eqnarray}
 Here
 \begin{eqnarray}
F'_{0}=\frac{F^{2}_{0}}{F^{2}_{0}+F^{2}_{1}+F^{2}_{2}+F^{2}_{3}},\nonumber\\
F'_{1}=\frac{F^{2}_{1}}{F^{2}_{0}+F^{2}_{1}+F^{2}_{2}+F^{2}_{3}},\nonumber\\
F'_{2}=\frac{F^{2}_{2}}{F^{2}_{0}+F^{2}_{1}+F^{2}_{2}+F^{2}_{3}},\nonumber\\
F'_{3}=\frac{F^{2}_{3}}{F^{2}_{0}+F^{2}_{1}+F^{2}_{2}+F^{2}_{3}}.\label{newfidelity}
  \end{eqnarray}
 If $F_{0}$ satisfies the relation
  \begin{eqnarray}
 F_{0}>\frac{1}{4}[3-2F_{1}-2F_{2}-\sqrt{1+4(F_{1}+F_{2})-12(F^{2}_{1}+F^{2}_{2})-8F_{1}F_{2}}],
  \end{eqnarray}
 they can obtain $F'_{0}>F_{0}$.
\section{MEPP for multipartite atomic entangled state for phase-flip error}
In above section, we briefly explained the EPP for a bit-flip error. The phase-flip error also exists
in a practical noisy environment. The phase-flip error cannot
purified directly. It is usually converted into a bit-flip error and purified in a next round.
For example, a bit-flip error will make the $|\Phi^{+}_{0}\rangle$ become $|\Phi^{-}_{0}\rangle$ with
a probability of $1-F$.  The mixed state  is written as
\begin{eqnarray}
\rho_{P}=F|\Phi^{+}_{0}\rangle\langle\Phi^{+}_{0}|+(1-F)|\Phi^{-}_{0}\rangle\langle\Phi^{-}_{0}|.\label{mixed4}
\end{eqnarray}
 Before  starting  to this MEPP, they first perform the Hadamard
 operations on their atoms and make
 \begin{eqnarray}
 |\Phi^{+}_{0}\rangle\rightarrow |\Psi^{+}_{0}\rangle=\frac{1}{2}(|g_{L}\rangle|g_{L}\rangle|g_{L}\rangle+|g_{L}\rangle|g_{R}\rangle|g_{R}\rangle+|g_{R}\rangle|g_{L}\rangle|g_{R}\rangle+|g_{R}\rangle|g_{R}\rangle|g_{L}\rangle),\nonumber\\
  |\Phi^{-}_{0}\rangle\rightarrow |\Psi^{-}_{0}\rangle=\frac{1}{2}(|g_{L}\rangle|g_{L}\rangle|g_{R}\rangle+|g_{L}\rangle|g_{R}\rangle|g_{L}\rangle+|g_{R}\rangle|g_{L}\rangle|g_{L}\rangle+|g_{R}\rangle|g_{R}\rangle|g_{R}\rangle).\nonumber\\\label{phase}
 \end{eqnarray}
 So the state in Eq. (\ref{mixed4}) becomes
 \begin{eqnarray}
\rho_{T}=F|\Psi^{+}_{0}\rangle\langle\Psi^{+}_{0}|+(1-F)|\Psi^{-}_{0}\rangle\langle\Psi^{-}_{0}|.\label{mixed5}
\end{eqnarray}
With the same principle as the bit-flip error correction, the two pairs of mixed states can be viewed as the combinations of four pure states:
With the probability of $F^{2}$, it is in the state $|\Psi^{+}_{0}\rangle_{a_{1}b_{1}c_{1}}\otimes|\Psi^{+}_{0}\rangle_{a_{2}b_{2}c_{2}}$, with the
equal probability $F(1-F)$, it is in the state $|\Psi^{+}_{0}\rangle_{a_{1}b_{1}c_{1}}\otimes|\Psi^{-}_{0}\rangle_{a_{2}b_{2}c_{2}}$ or
$|\Psi^{-}_{0}\rangle_{a_{1}b_{1}c_{1}}\otimes|\Psi^{+}_{0}\rangle_{a_{2}b_{2}c_{2}}$, with the probability of $(1-F)^{2}$, it is in the state
$|\Psi^{-}_{0}\rangle_{a_{1}b_{1}c_{1}}\otimes|\Psi^{-}_{0}\rangle_{a_{2}b_{2}c_{2}}$.

We first discuss the case of the cross-combination item $|\Psi^{+}_{0}\rangle_{a_{1}b_{1}c_{1}}\otimes|\Psi^{-}_{0}\rangle_{a_{2}b_{2}c_{2}}$.
It can be written as
 \begin{eqnarray}
 &&|\Psi^{+}_{0}\rangle_{a_{1}b_{1}c_{1}}\otimes|\Psi^{-}_{0}\rangle_{a_{2}b_{2}c_{2}}\nonumber\\
 &=&\frac{1}{2}(|g_{L}\rangle_{a_{1}}|g_{L}\rangle_{b_{1}}|g_{L}\rangle_{c_{1}}+|g_{L}\rangle_{a_{1}}|g_{R}\rangle_{b_{1}}|g_{R}\rangle_{c_{1}}\nonumber\\
 &+&|g_{R}\rangle_{a_{1}}|g_{L}\rangle_{b_{1}}|g_{R}\rangle_{c_{1}}+|g_{R}\rangle_{a_{1}}|g_{R}\rangle_{b_{1}}|g_{L}\rangle_{c_{1}})\nonumber\\
 &\otimes&\frac{1}{2}(|g_{L}\rangle_{a_{2}}|g_{L}\rangle_{b_{2}}|g_{R}\rangle_{c_{2}}+|g_{L}\rangle_{a_{2}}|g_{R}\rangle_{b_{2}}|g_{L}\rangle_{c_{2}}\nonumber\\
 &+&|g_{R}\rangle_{a_{2}}|g_{L}\rangle_{b_{2}}|g_{L}\rangle_{c_{2}}+|g_{R}\rangle_{a_{2}}|g_{R}\rangle_{b_{2}}|g_{R}\rangle_{c_{2}})\nonumber\\
 &=&\frac{1}{4}(|g_{L}\rangle_{a_{1}}|g_{L}\rangle_{b_{1}}|g_{L}\rangle_{c_{1}}|g_{L}\rangle_{a_{2}}|g_{L}\rangle_{b_{2}}|g_{R}\rangle_{c_{2}}+|g_{L}\rangle_{a_{1}}|g_{L}\rangle_{b_{1}}|g_{L}\rangle_{c_{1}}|g_{L}\rangle_{a_{2}}|g_{R}\rangle_{b_{2}}|g_{L}\rangle_{c_{2}}\nonumber\\
&+&|g_{L}\rangle_{a_{1}}|g_{L}\rangle_{b_{1}}|g_{L}\rangle_{c_{1}}|g_{R}\rangle_{a_{2}}|g_{L}\rangle_{b_{2}}|g_{L}\rangle_{c_{2}}+|g_{L}\rangle_{a_{1}}|g_{L}\rangle_{b_{1}}|g_{L}\rangle_{c_{1}}|g_{R}\rangle_{a_{2}}|g_{R}\rangle_{b_{2}}|g_{R}\rangle_{c_{2}}\nonumber\\
&+&|g_{L}\rangle_{a_{1}}|g_{R}\rangle_{b_{1}}|g_{R}\rangle_{c_{1}}|g_{L}\rangle_{a_{2}}|g_{L}\rangle_{b_{2}}|g_{R}\rangle_{c_{2}}+|g_{L}\rangle_{a_{1}}|g_{R}\rangle_{b_{1}}|g_{R}\rangle_{c_{1}}|g_{L}\rangle_{a_{2}}|g_{R}\rangle_{b_{2}}|g_{L}\rangle_{c_{2}}\nonumber\\
&+&|g_{L}\rangle_{a_{1}}|g_{R}\rangle_{b_{1}}|g_{R}\rangle_{c_{1}}|g_{R}\rangle_{a_{2}}|g_{L}\rangle_{b_{2}}|g_{L}\rangle_{c_{2}}+|g_{L}\rangle_{a_{1}}|g_{R}\rangle_{b_{1}}|g_{R}\rangle_{c_{1}}|g_{R}\rangle_{a_{2}}|g_{R}\rangle_{b_{2}}|g_{R}\rangle_{c_{2}}\nonumber\\
&+&|g_{R}\rangle_{a_{1}}|g_{L}\rangle_{b_{1}}|g_{R}\rangle_{c_{1}}|g_{L}\rangle_{a_{2}}|g_{L}\rangle_{b_{2}}|g_{R}\rangle_{c_{2}}+|g_{R}\rangle_{a_{1}}|g_{L}\rangle_{b_{1}}|g_{R}\rangle_{c_{1}}|g_{L}\rangle_{a_{2}}|g_{R}\rangle_{b_{2}}|g_{L}\rangle_{c_{2}}\nonumber\\
&+&|g_{R}\rangle_{a_{1}}|g_{L}\rangle_{b_{1}}|g_{R}\rangle_{c_{1}}|g_{R}\rangle_{a_{2}}|g_{L}\rangle_{b_{2}}|g_{L}\rangle_{c_{2}}+|g_{R}\rangle_{a_{1}}|g_{L}\rangle_{b_{1}}|g_{R}\rangle_{c_{1}}|g_{R}\rangle_{a_{2}}|g_{R}\rangle_{b_{2}}|g_{R}\rangle_{c_{2}}\nonumber\\
&+&|g_{R}\rangle_{a_{1}}|g_{R}\rangle_{b_{1}}|g_{L}\rangle_{c_{1}}|g_{L}\rangle_{a_{2}}|g_{L}\rangle_{b_{2}}|g_{R}\rangle_{c_{2}}+|g_{R}\rangle_{a_{1}}|g_{R}\rangle_{b_{1}}|g_{L}\rangle_{c_{1}}|g_{L}\rangle_{a_{2}}|g_{R}\rangle_{b_{2}}|g_{L}\rangle_{c_{2}}\nonumber\\
&+&|g_{R}\rangle_{a_{1}}|g_{R}\rangle_{b_{1}}|g_{L}\rangle_{c_{1}}|g_{R}\rangle_{a_{2}}|g_{L}\rangle_{b_{2}}|g_{L}\rangle_{c_{2}}+|g_{R}\rangle_{a_{1}}|g_{R}\rangle_{b_{1}}|g_{L}\rangle_{c_{1}}|g_{R}\rangle_{a_{2}}|g_{R}\rangle_{b_{2}}|g_{R}\rangle_{c_{2}}).\nonumber\\\label{cross5}
 \end{eqnarray}
From Eq. (\ref{cross5}), if Alice, Bob and Charlie let their $\frac{1}{\sqrt{2}}(|L\rangle+|R\rangle)$ photons entrance the cavities, respectively, they cannot lead the
cases that all the polarization of the photons be flipped. The another cross-combination $|\Psi^{-}_{0}\rangle_{a_{1}b_{1}c_{1}}\otimes|\Psi^{+}_{0}\rangle_{a_{2}b_{2}c_{2}}$ also cannot lead all the polarization of the photons be flipped.
Now we discuss the case of  $|\Psi^{+}_{0}\rangle_{a_{1}b_{1}c_{1}}\otimes|\Psi^{+}_{0}\rangle_{a_{2}b_{2}c_{2}}$. It can be written as
 \begin{eqnarray}
 &&|\Psi^{+}_{0}\rangle_{a_{1}b_{1}c_{1}}\otimes|\Psi^{+}_{0}\rangle_{a_{2}b_{2}c_{2}}\nonumber\\
  &=&\frac{1}{2}(|g_{L}\rangle_{a_{1}}|g_{L}\rangle_{b_{1}}|g_{L}\rangle_{c_{1}}+|g_{L}\rangle_{a_{1}}|g_{R}\rangle_{b_{1}}|g_{R}\rangle_{c_{1}}\nonumber\\
  &+&|g_{R}\rangle_{a_{1}}|g_{L}\rangle_{b_{1}}|g_{R}\rangle_{c_{1}}+|g_{R}\rangle_{a_{1}}|g_{R}\rangle_{b_{1}}|g_{L}\rangle_{c_{1}})\nonumber\\
 &\otimes&\frac{1}{2}(|g_{L}\rangle_{a_{2}}|g_{L}\rangle_{b_{2}}|g_{L}\rangle_{c_{2}}+|g_{L}\rangle_{a_{2}}|g_{R}\rangle_{b_{2}}|g_{R}\rangle_{c_{2}}\nonumber\\
 &+&|g_{R}\rangle_{a_{2}}|g_{L}\rangle_{b_{2}}|g_{R}\rangle_{c_{2}}+|g_{R}\rangle_{a_{2}}|g_{R}\rangle_{b_{2}}|g_{L}\rangle_{c_{2}})\nonumber\\
 &=&\frac{1}{4}(|g_{L}\rangle_{a_{1}}|g_{L}\rangle_{b_{1}}|g_{L}\rangle_{c_{1}}|g_{L}\rangle_{a_{2}}|g_{L}\rangle_{b_{2}}|g_{L}\rangle_{c_{2}}+|g_{L}\rangle_{a_{1}}|g_{L}\rangle_{b_{1}}|g_{L}\rangle_{c_{1}}|g_{L}\rangle_{a_{2}}|g_{R}\rangle_{b_{2}}|g_{R}\rangle_{c_{2}}\nonumber\\
 &+&|g_{L}\rangle_{a_{1}}|g_{L}\rangle_{b_{1}}|g_{L}\rangle_{c_{1}}|g_{R}\rangle_{a_{2}}|g_{L}\rangle_{b_{2}}|g_{R}\rangle_{c_{2}}+|g_{L}\rangle_{a_{1}}|g_{L}\rangle_{b_{1}}|g_{L}\rangle_{c_{1}}|g_{R}\rangle_{a_{2}}|g_{R}\rangle_{b_{2}}|g_{L}\rangle_{c_{2}}\nonumber\\
 &+&|g_{L}\rangle_{a_{1}}|g_{R}\rangle_{b_{1}}|g_{R}\rangle_{c_{1}}|g_{L}\rangle_{a_{2}}|g_{L}\rangle_{b_{2}}|g_{L}\rangle_{c_{2}}+|g_{L}\rangle_{a_{1}}|g_{R}\rangle_{b_{1}}|g_{R}\rangle_{c_{1}}|g_{L}\rangle_{a_{2}}|g_{R}\rangle_{b_{2}}|g_{R}\rangle_{c_{2}}\nonumber\\
 &+&|g_{L}\rangle_{a_{1}}|g_{R}\rangle_{b_{1}}|g_{R}\rangle_{c_{1}}|g_{R}\rangle_{a_{2}}|g_{L}\rangle_{b_{2}}|g_{R}\rangle_{c_{2}}+|g_{L}\rangle_{a_{1}}|g_{R}\rangle_{b_{1}}|g_{R}\rangle_{c_{1}}|g_{R}\rangle_{a_{2}}|g_{R}\rangle_{b_{2}}|g_{L}\rangle_{c_{2}}\nonumber\\
 &+&|g_{R}\rangle_{a_{1}}|g_{L}\rangle_{b_{1}}|g_{R}\rangle_{c_{1}}|g_{L}\rangle_{a_{2}}|g_{L}\rangle_{b_{2}}|g_{L}\rangle_{c_{2}}+|g_{R}\rangle_{a_{1}}|g_{L}\rangle_{b_{1}}|g_{R}\rangle_{c_{1}}|g_{L}\rangle_{a_{2}}|g_{R}\rangle_{b_{2}}|g_{R}\rangle_{c_{2}}\nonumber\\
 &+&|g_{R}\rangle_{a_{1}}|g_{L}\rangle_{b_{1}}|g_{R}\rangle_{c_{1}}|g_{R}\rangle_{a_{2}}|g_{L}\rangle_{b_{2}}|g_{R}\rangle_{c_{2}}+|g_{R}\rangle_{a_{1}}|g_{L}\rangle_{b_{1}}|g_{R}\rangle_{c_{1}}|g_{R}\rangle_{a_{2}}|g_{R}\rangle_{b_{2}}|g_{L}\rangle_{c_{2}}\nonumber\\
 &+&|g_{R}\rangle_{a_{1}}|g_{R}\rangle_{b_{1}}|g_{L}\rangle_{c_{1}}|g_{L}\rangle_{a_{2}}|g_{L}\rangle_{b_{2}}|g_{L}\rangle_{c_{2}}+|g_{R}\rangle_{a_{1}}|g_{R}\rangle_{b_{1}}|g_{L}\rangle_{c_{1}}|g_{L}\rangle_{a_{2}}|g_{R}\rangle_{b_{2}}|g_{R}\rangle_{c_{2}}\nonumber\\
 &+&|g_{R}\rangle_{a_{1}}|g_{R}\rangle_{b_{1}}|g_{L}\rangle_{c_{1}}|g_{R}\rangle_{a_{2}}|g_{L}\rangle_{b_{2}}|g_{R}\rangle_{c_{2}}+|g_{R}\rangle_{a_{1}}|g_{R}\rangle_{b_{1}}|g_{L}\rangle_{c_{1}}|g_{R}\rangle_{a_{2}}|g_{R}\rangle_{b_{2}}|g_{L}\rangle_{c_{2}}.\nonumber\\\label{cross6}
 \end{eqnarray}
From Eq. (\ref{cross6}), if Alice, Bob and Charlie let their $\frac{1}{\sqrt{2}}(|L\rangle+|R\rangle)$ photons entrance the cavities and choose the case that
all the photons are flipped, they will obtain
 \begin{eqnarray}
 |\psi\rangle&=&\frac{1}{2}(|g_{L}\rangle_{a_{1}}|g_{L}\rangle_{b_{1}}|g_{L}\rangle_{c_{1}}|g_{L}\rangle_{a_{2}}|g_{L}\rangle_{b_{2}}|g_{L}\rangle_{c_{2}}+|g_{L}\rangle_{a_{1}}|g_{R}\rangle_{b_{1}}|g_{R}\rangle_{c_{1}}|g_{L}\rangle_{a_{2}}|g_{R}\rangle_{b_{2}}|g_{R}\rangle_{c_{2}}\nonumber\\
 &+&|g_{R}\rangle_{a_{1}}|g_{L}\rangle_{b_{1}}|g_{R}\rangle_{c_{1}}|g_{R}\rangle_{a_{2}}|g_{L}\rangle_{b_{2}}|g_{R}\rangle_{c_{2}}+|g_{R}\rangle_{a_{1}}|g_{R}\rangle_{b_{1}}|g_{L}\rangle_{c_{1}}|g_{R}\rangle_{a_{2}}|g_{R}\rangle_{b_{2}}|g_{L}\rangle_{c_{2}}).\nonumber\\
 \end{eqnarray}

 The item $|\Psi^{-}_{0}\rangle_{a_{1}b_{1}c_{1}}\otimes|\Psi^{-}_{0}\rangle_{a_{2}b_{2}c_{2}}$ can also lead all the photons be flipped and it  becomes
   \begin{eqnarray}
    |\psi\rangle'&=&\frac{1}{2}(|g_{L}\rangle_{a_{1}}|g_{L}\rangle_{b_{1}}|g_{R}\rangle_{c_{1}}|g_{L}\rangle_{a_{2}}|g_{L}\rangle_{b_{2}}|g_{R}\rangle_{c_{2}}+|g_{L}\rangle_{a_{1}}|g_{R}\rangle_{b_{1}}|g_{L}\rangle_{c_{1}}|g_{L}\rangle_{a_{2}}|g_{R}\rangle_{b_{2}}|g_{L}\rangle_{c_{2}}\nonumber\\
 &+&|g_{R}\rangle_{a_{1}}|g_{L}\rangle_{b_{1}}|g_{L}\rangle_{c_{1}}|g_{R}\rangle_{a_{2}}|g_{L}\rangle_{b_{2}}|g_{L}\rangle_{c_{2}}+|g_{R}\rangle_{a_{1}}|g_{R}\rangle_{b_{1}}|g_{R}\rangle_{c_{1}}|g_{R}\rangle_{a_{2}}|g_{R}\rangle_{b_{2}}|g_{R}\rangle_{c_{2}}).\nonumber\\
   \end{eqnarray}

   Then all the three parties perform  Hadamard operations on the $a_{2}$, $b_{2}$ and $c_{2}$ atoms  and measure them respectively.
   If the measurement results are $|g_{L}\rangle_{a_{2}}|g_{L}\rangle_{b_{2}}|g_{L}\rangle_{c_{2}}$, $|g_{L}\rangle_{a_{2}}|g_{R}\rangle_{b_{2}}|g_{R}\rangle_{c_{2}}$,
   $|g_{R}\rangle_{a_{2}}|g_{L}\rangle_{b_{2}}|g_{R}\rangle_{c_{2}}$, or $|g_{R}\rangle_{a_{2}}|g_{R}\rangle_{b_{2}}|g_{L}\rangle_{c_{2}}$, the $ |\psi\rangle$ becomes
   \begin{eqnarray}
   |\psi\rangle&\rightarrow&\frac{1}{2}(|g_{L}\rangle_{a_{1}}|g_{L}\rangle_{b_{1}}|g_{L}\rangle_{c_{1}}+|g_{L}\rangle_{a_{1}}|g_{R}\rangle_{b_{1}}|g_{R}\rangle_{c_{1}}\nonumber\\
   &+&|g_{R}\rangle_{a_{1}}|g_{L}\rangle_{b_{1}}|g_{R}\rangle_{c_{1}}+|g_{R}\rangle_{a_{1}}|g_{R}\rangle_{b_{1}}|g_{L}\rangle_{c_{1}},\label{collapse1}
    \end{eqnarray}
    the $ |\psi\rangle'$ becomes
    \begin{eqnarray}
  |\psi\rangle'&\rightarrow&\frac{1}{2}(|g_{L}\rangle_{a_{1}}|g_{L}\rangle_{b_{1}}|g_{R}\rangle_{c_{1}}+|g_{L}\rangle_{a_{1}}|g_{R}\rangle_{b_{1}}|g_{L}\rangle_{c_{1}}\nonumber\\
  &+&|g_{R}\rangle_{a_{1}}|g_{L}\rangle_{b_{1}}|g_{L}\rangle_{c_{1}}+|g_{R}\rangle_{a_{1}}|g_{R}\rangle_{b_{1}}|g_{R}\rangle_{c_{1}}.\label{collapse2} \end{eqnarray}
  Interestingly, Eqs. (\ref{collapse1}) and (\ref{collapse2}) have the same form of Eq. (\ref{phase}). So they only need to perform  Hadamard operations
  on their atoms to recover the $|\Psi^{+}_{0}\rangle$ and $ |\Psi^{-}_{0}\rangle$ to $|\Phi^{+}_{0}\rangle$ and $|\Phi^{-}_{0}\rangle$, respectively and obtain
  a new mixed state
   \begin{eqnarray}
\rho'_{P}=F'|\Phi^{+}_{0}\rangle\langle\Phi^{+}_{0}|+(1-F')|\Phi^{-}_{0}\rangle\langle\Phi^{-}_{0}|.\label{mixed6}
\end{eqnarray}
On the other hand, if the measurement results performed on the atoms are  $|g_{L}\rangle_{a_{2}}|g_{L}\rangle_{b_{2}}|g_{R}\rangle_{c_{2}}$, $|g_{L}\rangle_{a_{2}}|g_{R}\rangle_{b_{2}}|g_{L}\rangle_{c_{2}}$,
   $|g_{R}\rangle_{a_{2}}|g_{L}\rangle_{b_{2}}|g_{L}\rangle_{c_{2}}$, or $|g_{R}\rangle_{a_{2}}|g_{R}\rangle_{b_{2}}|g_{R}\rangle_{c_{2}}$, by performing a phase-flip operation on one of the atoms on $a_{1}$, $b_{1}$ or $c_{1}$, they can obtain the same high quality mixed state as shown in Eq. (\ref{mixed6}).
By choosing the cases that all the photons are flipped and detected by D$_{2}$D$_{4}$D$_{6}$, they can ultimately obtain the new fidelity $F'$ of the atomic mixed state
with $F'=\frac{F^{2}}{F^{2}+(1-F)^{2}}$.
Certainly, if a mixed state contains all the phase-flip error and can be written as
\begin{eqnarray}
   \rho_{P}''=F_{0}|\Phi^{+}_{0}\rangle\langle\Phi^{-}_{0}|+F_{1}|\Phi^{-}_{1}\rangle\langle\Phi^{-}_{1}|
   +F_{2}|\Phi^{-}_{2}\rangle\langle\Phi^{-}_{2}|+|F_{3}|\Phi^{-}_{3}\rangle\langle\Phi^{-}_{3}|.\label{mixed7}
 \end{eqnarray}
After correcting the phase-flip error, one can also obtain a new mixed state with the same fidelity as shown in Eq. (\ref{newfidelity}).
Essentially, in Eq. (\ref{mixed7}), it  also contains  the bit-flip errors. Therefore, during the whole MEPP,
 both the bit-flip and phase-flip errors correction are all required. In a practical operation, they can first perform the
 bit-flip error correction and then perform the phase-flip error correction, or first perform the phase-flip error correction, and
 then perform the bit-flip error correction, which is similar as Ref. \cite{murao}. In this way, one can purify an arbitrary
 three-atom mixed entangled state.

\section{MEPP for N-particle GHZ state}
In the previous sections, we have explained the MEPP for the case of three-particle system.
 This protocol can also be extended to $N$-partite mixed entangled systems with both bit-flip
and phase-flip errors.
Suppose  the $N$-atom entangled state is prepared in one of the GHZ state
\begin{eqnarray}
|\Phi^{+}_{0}\rangle_{N}=\frac{1}{\sqrt{2}}(|g_{L}\rangle_{a_{1}}|g_{L}\rangle_{b_{1}}\cdots|g_{L}\rangle_{n_{1}}+|g_{R}\rangle_{a_{1}}|g_{R}\rangle_{b_{1}}\cdots|g_{R}\rangle_{n_{1}}).\label{Nparticle}
\end{eqnarray}
The $N$ atoms are belonged to Alice, Bob, Charlie, $\cdots$, Kite, etc, as shown in Fig. 2.
The noise will lead it have a bit-flip error can become
\begin{eqnarray}
|\Phi^{+}_{1}\rangle_{N}=\frac{1}{\sqrt{2}}(|g_{R}\rangle_{a_{1}}|g_{L}\rangle_{b_{1}}\cdots|g_{L}\rangle_{n_{1}}+|g_{L}\rangle_{a_{1}}|g_{R}\rangle_{b_{1}}\cdots|g_{R}\rangle_{n_{1}}).\label{Nparticle1}
\end{eqnarray}
So the mixed state can be written as
\begin{eqnarray}
\rho_{N}=F|\Phi^{+}_{0}\rangle_{N}\langle\Phi^{+}_{0}|+(1-F)F|\Phi^{+}_{1}\rangle_{N}\langle\Phi^{+}_{1}|.\label{Nmixed}
\end{eqnarray}
As shown in Fig. 2, after the $N$ photons passing through the cavities, respectively, they only to choose the cases that all the photons are
have the same polarization. Then the $2N$ atoms will collapse to
\begin{eqnarray}
|\Phi^{+}_{0}\rangle_{2N}&=&\frac{1}{\sqrt{2}}(|g_{L}\rangle_{a_{1}}|g_{L}\rangle_{b_{1}}\cdots|g_{L}\rangle_{n_{1}}|g_{L}\rangle_{a_{2}}|g_{L}\rangle_{b_{2}}|g_{L}\rangle_{n_{2}}\nonumber\\
&+&|g_{R}\rangle_{a_{1}}|g_{R}\rangle_{b_{1}}\cdots|g_{R}\rangle_{n_{1}}|g_{R}\rangle_{a_{2}}|g_{R}\rangle_{b_{2}}\cdots|g_{R}\rangle_{n_{2}}),\label{2Nparticle}
\end{eqnarray}
 with the probability of $\frac{F^{2}}{2}$.
 It also will collapse to
 \begin{eqnarray}
|\Phi^{+}_{1}\rangle_{2N}&=&\frac{1}{\sqrt{2}}(|g_{R}\rangle_{a_{1}}|g_{L}\rangle_{b_{1}}\cdots|g_{L}\rangle_{n_{1}}|g_{R}\rangle_{a_{2}}|g_{L}\rangle_{b_{2}}|g_{L}\rangle_{n_{2}}\nonumber\\
&+&|g_{L}\rangle_{a_{1}}|g_{R}\rangle_{b_{1}}\cdots|g_{R}\rangle_{n_{1}}|g_{L}\rangle_{a_{2}}|g_{R}\rangle_{b_{2}}\cdots|g_{R}\rangle_{n_{2}}),\label{2Nparticle2}
\end{eqnarray}
 with the probability of $\frac{(1-F)^{2}}{2}$. Finally, by performing the Hadamard operations on each atoms $a_{2}$, $b_{2}$, $\cdots$, $n_{2}$,
 and measuring them in the basis $\{|g_{L}\rangle,|g_{R}\rangle\}$, they will ultimately obtain the $N$-atom high fidelity mixed  state, with
 $F'=\frac{F^{2}}{F^{2}+(1-F)^{2}}$. If the bit-flip error occurs on the other qubits, it can also be purified in the same principle.
 On the other hand, if the phase-flip errors, they first perform the Hadamard operations on each atoms to convert it to the bit-flip error and
 purify it in the next step. So both bit-flip and phase-flip error can be purified.
 In this way, one can purify an arbitrary $N$-particle mixed entangled state.
 \section{Discussion and summary}
 So far, we have fully described our MEPP. In this MEPP, we resort some single photons to complete the task.
 From Eq. (\ref{relation1}), the polarization of the photon is flipped or not flipped are decided by the
 state of  two atoms. The photonic Faraday rotation essentially acts as the role of CNOT gate between the photon
 and atoms. Moreover, according to the polarization of the photon, one can check the parity of the two atoms. From above
 description, if the initial state of photon is $\frac{1}{\sqrt{2}}(|L\rangle+|R\rangle)$, it will be flipped if the atoms are in the
 odd parity states $|g_{R}\rangle|g_{L}\rangle$ or $|g_{L}\rangle|g_{R}\rangle$, while it does not change if the atoms are in the
 even parity states  $|g_{L}\rangle|g_{L}\rangle$ or $|g_{R}\rangle|g_{R}\rangle$. Interestingly, during the whole process, we do not need to
 measure the atoms directly, which lead this MEPP essentially is the quantum nondemolition (QND) like the cross-Kerr nonlinearity \cite{shengpra1}.
Actually, in the early works of atomic entanglement purification, Gonta and van Loock have discussed a different dynamical entanglement purification scheme\cite{purificationcavity}. They can purify dynamically a bipartite entangled
state using short chains of atoms coupled to high-finesse optical cavities. When the atomic entangled  pairs pass through the cavities,
the atomic
evolution can be mediated by the cavity field with the effective Hamiltonian described in the Heisenberg XY
model. On the other hand, Reichle \emph{et al.} have reported their experimental  work about purification of two-atom entanglement \cite{purificationatom}.
In their experiment, they should  confine four $^{9}$Be$^{+}$ ions in one
trapping zone of a linear multi-zone Paul trap, which lead it not suitable for long-distance quantum communication and distributed quantum computation. In a practical experiment, the $^{87}$Rb atom trapped in a fiber-based Fabry-Perot cavity is a good candidate. As shown
in Ref.\cite{atomexperiemnt}, the generated ground states $|g_{L}\rangle$ and $|g_{R}\rangle$ can be the states $|F=2,m_{F}=\pm1\rangle$ of the level
$5S_{\frac{1}{2}}$, respectively. The excited state $|e\rangle$ can be the state  $|F'=3,m_{F}=0\rangle$ of the level $5S_{\frac{3}{2}}$.
So the corresponding transition frequency at $\lambda=780$ nm ($D_{2}$ line) is $\omega_{0}=2\pi c/\lambda$.
The cavity decay rate $\kappa=2\pi\times53$ MHz with relevant Q factor $Q=\omega_{c}/(2\kappa)=3.63\times10^{6}$. The length of the cavity is
38.6$\mu$m.

During the whole protocol, we only discuss the case that the virtual excitations of the atoms and consider low-Q cavities, fibers without absorption,
ideal single photon detectors and perfect measurements on atoms. As shown in Ref. \cite{imperfection}, the coupling and transmission of each photon
through the single-mode optical fiber is $T_{f}=0.2$, and the single-photon detector efficiency is $\eta_{d}=0.28$ \cite{detectorefficency}. The transmission efficiency of each
photon through the other optical elements is $\eta_{0}=0.95$. Suppose the  measurement efficiency of the atom is $\eta_{a}=0.95$ \cite{atommeasurement}. We can estimate
the success probability of this MEPP for a bit-flip error correction as $P=P_{p}\times T_{f}\times \eta_{0}\times\eta^{N}_{d}\times\eta^{N}_{a}$, with $P_{p}=F^{2}+(1-F)^{2}$.
$N$ is the number of the atoms. If the initial fidelity $F=0.8$, and $N=3$, we can calculate $P\simeq2.43\times10^{-3}$.
Obviously, the single-photon detector efficiency greatly affect the total success probability because all the photons will be detected.

In conclusion, we have presented an MEPP for purifying  multipartite atomic entanglement using photonic Faraday rotation. Through a single-photon input-output process in
each cavity-QED, one can obtain the higher fidelity atomic mixed state from two pair of low fidelity mixed states with some success probability.
During the whole protocol, we resort to the low-Q cavities, some linear optical elements, the three-level atoms and the single-photon detectors
to complete the task, which lead this MEPP be suitable for current experimental condition.

\section*{ACKNOWLEDGEMENTS}
This work was supported by the National Natural Science Foundation
of China under Grant No. 11104159,   Open Research
Fund Program of the State Key Laboratory of
Low-Dimensional Quantum Physics Scientific, Tsinghua University, Open Research Fund Program of National Laboratory of Solid State Microstructures under Grant No. M25020 and M25022, and the Project
Funded by the Priority Academic Program Development of Jiangsu
Higher Education Institutions.

\end{document}